\documentclass[twocolumn,showpacs,preprintnumbers,amsmath,amssymb]{revtex4}

\usepackage{graphicx}
\usepackage{dcolumn}
\usepackage{bm}

\begin{document}


\title{Demonstration of Controllable Temporal Distinguishability in a Three-Photon State}

\author{B. H. Liu$^{1\dag}$, F. W. Sun$^{1\dag}$, Y. X. Gong$^1$, Y. F. Huang$^1$, Z. Y. Ou$^{1,2,*}$, and G. C. Guo$^1$}
 \affiliation{$^1$Key Laboratory of Quantum Information,
 University of Science and Technology of China, \\CAS, Hefei, 230026, the People's Republic of China
 \\$^2$Department of Physics, Indiana
University-Purdue University Indianapolis, 402 N. Blackford
Street, Indianapolis, IN 46202, USA\\
$^{\dag}$These two authors contribute equally. $^*$Corresponding
author: zou@iupui.edu}

\date{\today}

\begin{abstract}
Multi-photon interference is at the heart of the recently proposed
linear optical quantum computing scheme\cite{KLM} and plays an
essential role in many protocols in quantum
information\cite{bou,pan,zhao}. Indistinguishability is what leads
to the effect of quantum interference. Optical interferometers
such as Michaelson interferometer provide a measure for
second-order coherence at one-photon level\cite{wolf} and
Hong-Ou-Mandel interferometer\cite{hom} was widely employed to
describe two-photon entanglement and
indistinguishability\cite{wam,ser,wong}. However, there is not an
effective way for a system of more than two photons. Recently, a
new interferometric scheme \cite{sun1,res,sun2} was
proposed\cite{ou} to quantify the degree of multi-photon
distinguishability. Here we report an experiment to implement the
scheme for three-photon case. We are able to generate three
photons with different degrees of temporal distinguishability and
demonstrate how to characterize them by the visibility of
three-photon interference. This method of quantitative description
of multi-photon indistinguishability will have practical
implications in the implementation of quantum information
protocols.
\end{abstract}

\maketitle

Early pioneers of quantum optics developed a complete theory of
quantum coherence\cite{gla,sud}, which was able to explain and
predict a number of quantum phenomena of light such as photon
anti-bunching, sub-Poissonian photon statistics, and squeezed
state of light\cite{w-m}. However, the theory was geared in close
connection to the classical coherence theory with an emphasis on
the wave aspect of light and is best to characterize coherence in
the second-order of the field amplitudes. With the recent advent
of quantum information science, most of the applications are
photon-number based system, i.e., a system with a definite number
of photons. Thus the Glauber's quantum coherence theory fell short
to give a direct account of the quantum entanglement of a
multi-photon system.

Quantum entanglement is best described by the quantum interference
effect. As is generally believed, indistinguishability of photon's
paths directly leads to the quantum interference effect. Early
method\cite{wam,ser,wong} for characterizing the two-photon
temporal distinguishability is by sending the fields into a
Hong-Ou-Mandel two-photon interferometer\cite{hom} and measuring
the visibility of the interference dip. A number of
attempts\cite{orw1,tsu,ou2} were made to characterize the temporal
distinguishability for the two independent pairs of photons from
parametric down-conversion and a quantity ${\cal E/A}$ is
identified\cite{orw2,gis} to characterize the temporal
distinguishability between different pairs of down-converted
photons (photons within a pair are in the same temporal mode and
are indistinguishable under certain condition). However, a
generalization to arbitrary photon number is not possible until a
new multi-photon interferometric scheme, the so-called "NOON"
state projection measurement, was proposed\cite{sun1} and
realized\cite{res,sun2} recently. The new interferometric scheme
was initially used to demonstrate the N-photon de Broglie
wavelength\cite{wal,ste} without the need of a maximally entangled
N-photon state (the so-called "NOON" state)\cite{sun1,res,sun2}.
But it is shown\cite{ou} that the visibility or the relative depth
of the interference dip in the "NOON" state projection measurement
can be used to quantitatively characterize the different scenarios
of temporal distinguishability of a multi-photon system.

In this letter, we wish to report on an experimental procedure
based on the "NOON" state projection measurement scheme to
characterize the temporal distinguishability of a controllable
three-photon system generated from two parametric down-converters.
We observed a three-photon interference dip with 91\% visibility
when all three photons are indistinguishable and two dips with
respective 45\% and 39\% visibility when two among the three
photons become distinguishable. We compare the measured visibility
with a model of pulse pumped parametric down-conversion and obtain
good agreement.

\begin{figure}[htb]
\begin{center}
\includegraphics[width= 3in]{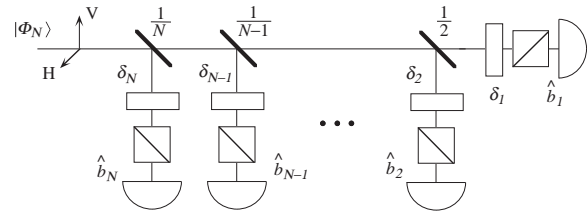}
\end{center}
\caption{\em A NOON-state projection measurement for N photons}
\label{fig1}
\end{figure}

We start by describing the "NOON" state projection measurement
shown in Fig.1 for N-photon case. The incoming field of both
horizontal(H) and vertical(V) polarizations is divided into N
equal parts by N-1 beam splitters. Each part passes through a wave
plate that introduces a relative phase shift of $\delta_k= 2\pi
(k-1)/N (k=1,...,N)$ between the H and V polarizations. It is then
projected to the 135$^{\circ}$-direction by a polarizer. The field
operator at the $k$th detector is expressed via the input
operators by
\begin{eqnarray}
\hat b_k = (\hat a_H - \hat a_V e^{i\delta_k})/\sqrt{2N}+\hat
b_{k0},~~~~(k= 1, ..., N) \label{bk}
\end{eqnarray}
where $\hat b_{k0}$ is related to the vacuum input to the beam
splitters. The N-photon detection rate is proportional to
\begin{eqnarray}
P_N = \bigg\langle \prod_{k=1}^{N}\hat b_k^{\dag}
\prod_{k=1}^{N}\hat b_k\bigg\rangle,\label{PN}
\end{eqnarray}
where the average is over the quantum state $|\Phi_N\rangle = \sum
c_k|k\rangle_k|N-k\rangle_V$ of $\hat a_{H,V}$. But because of the
identity
\begin{eqnarray}
\prod_{k=1}^{N}\hat b_k = (\hat a_H^N - \hat a_V^N)
/(2N)^{N/2},\label{bkiden}
\end{eqnarray}
Eq.(\ref{PN}) becomes
\begin{eqnarray}
P_N = 2 N!|\langle \Phi_N|NOON\rangle|^2/(2N)^N ,\label{PN1}
\end{eqnarray}
where $|NOON\rangle \equiv
(|N\rangle_H|0\rangle_V-|0\rangle_H|N\rangle_V)/\sqrt{2}$ is the
NOON-state\cite{ou3,dow}. So the N-photon coincidence rate is
proportional to the probability of projecting $|\Phi_N\rangle$ to
the NOON state.

If the input state $|\Phi_N\rangle$ has both non-zero $c_0$ and
$c_N$ and experiences a relative phase shift of $\delta$ between H
and V polarizations, it is easy to show that
\begin{eqnarray}
P_N \propto |c_0|^2 + |c_N|^2 -2 |c_0 c_N|\cos(N\delta +\delta_0)
,\label{PN2}
\end{eqnarray}
which demonstrates an N-photon de Broglie wave length. On the
other hand, if the input state is orthogonal to the NOON state, in
particular if $|\Phi_N\rangle = |k, N-k\rangle$ with $k\ne 0, N$,
we will have zero coincidence for N-photon detection, i.e., $P_N =
0$, due to orthogonal projection. When we examine
Eq.(\ref{bkiden}), we find this orthogonality stems from the
absence of the terms of $\hat a_H^k\hat a_V^{n-k} (k\ne 0,N)$. A
further examination shows that the disappearance of the
coefficients of $\hat a_H^k\hat a_V^{n-k} (k\ne 0,N)$ is because
of complete destructive N-photon interference\cite{hof}.

Of course, the above analysis is based on a single mode
description in which all N photons are in one indistinguishable
temporal mode. In reality, the N photons may not be in a single
temporal mode. Then we will not have complete destructive
interference and $P_N$ will be a non-zero value depending on the
degree of temporal distinguishability. This is precisely the
proposal by Ou\cite{ou} to use the visibility of the N-photon
interference for quantitative characterization of temporal
distinguishability of the $|k, N-k\rangle (k\ne 0, N)$ state. For
the state of $|N-1\rangle_H|1\rangle_V$, in particular, we have
the N-photon interference visibility ${\cal V}_N = m/(N-1)$ when
$m$ H-photons are in one temporal mode that is distinguishable
from other $N-m-1$ H-photons (the V-photon must have the same
temporal mode as the $m$ H-photons)\cite{ou}.

\begin{figure}[htb]
\begin{center}
\includegraphics[width= 3.2in]{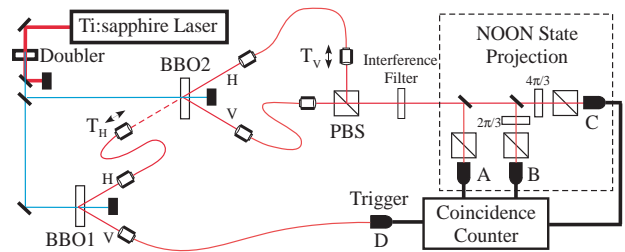}
\end{center}
\caption{\em Setup for the experiment.} \label{fig2}
\end{figure}

To demonstrate experimentally the procedure for characterizing
N-photon temporal distinguishability, we need to prepare a state
of the form $|N-1\rangle_H|1\rangle_V$ that is orthogonal to the
NOON state. Note that this state is fundamentally different from
the states used in Refs.\cite{res,sun2} where a non-orthogonal
state to the NOON state must be employed for the demonstration of
N-photon de Broglie wavelength. We are able to generate a
three-photon state in the form of $|2_H,1_V\rangle$ with
controllable temporal distinguishability by two type-II parametric
down-converters shown in Fig.2. Two BBO crystals are pumped
synchronously by two pulses of 150 fs length from a frequency
doubled Ti:sapphire laser operating at 780 nm. The H-photon from
BBO1 is injected into the H-polarization mode of BBO2 whereas the
V-photon of BBO1 is detected to produce a trigger for three-photon
coincidence. The H-photon from BBO2 together with the H-photon
from BBO1 are coupled into a single mode fiber and then are
combined by a polarization beam splitter (PBS) with the V-photon
from BBO2 via another single mode fiber. The combined fields
passes through an interference filter (IF) of 3 nm bandpass before
entering the "NOON" state projection measurement. The output of
the fiber coupler for the H-photons is mounted on a translation
stage for the adjustment of the relative delay $T_V$ between the
H- and V-photons. The relative delay $T_H$ between the two
H-photons can be adjusted by another translation stage on the
H-photon from BBO1. When $T_H >> T_c$ ($T_c$= the temporal width
of the photons determined by the bandpass of the IF), the two
H-photons are well separated and distinguishable but when $T_H <<
T_c$, the two H-photons becomes indistinguishable. The condition
$T_H << T_c$ can be found by blocking the V-photon and observing a
bump in two-photon coincidence of the two H-photons as we scan
$T_H$ (see the inset of Fig.3). Four-photon coincidence among ABCD
detectors as well as two-photon coincidence between any two of the
four detectors are measured. The four-photon coincidence is
equivalent to the three-photon coincidence of ABC gated on the
detection at D. The gated coincidence measurement ensures the two
H-photons come from different crystals for the controllable
distinguishability. The chance for two H-photons from the same
crystal is rare and is from higher order case of three pairs.

\begin{figure}[htb]
\begin{center}
\includegraphics[width= 3.2in]{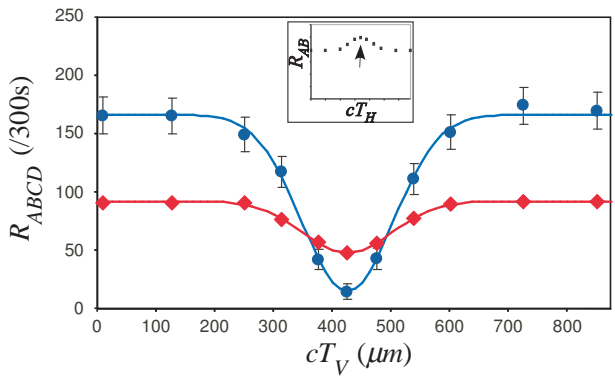}
\end{center}
\caption{\em Four-photon coincidence as a function of the relative
delay $cT_V$ between the V-photon and the H-photons for the case
of two overlapping H-photons ($T_H=0$). Circles (blue):
$R_{ABCD}$; Diamonds(red): $R_{ABCD}(2\times2)$ derived from
Eq.(\ref{2x2}).} \label{fig3}
\end{figure}

\begin{figure}[htb]
\begin{center}
\includegraphics[width= 3.2in]{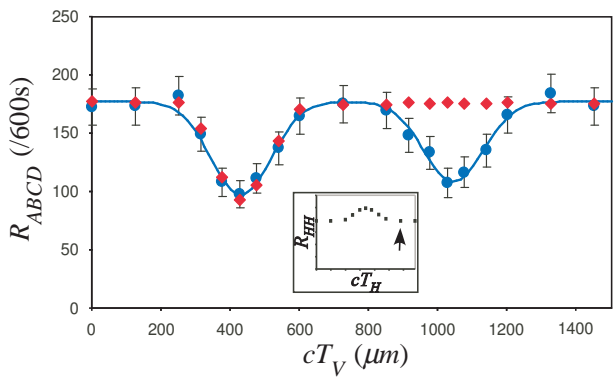}
\end{center}
\caption{\em Four-photon coincidence as a function of the relative
delay $cT_V$ between the V-photon and the H-photons for two well
separated H-photons ($T_H>>T_c$). Circles (blue): $R_{ABCD}$;
Diamonds(red): $R_{ABCD}(2\times2)$ derived from Eq.(\ref{2x2}).}
\label{fig4}
\end{figure}

The first experiment is performed when $T_H$ is set at zero. We
measure the four-photon coincidence among the ABCD detectors as we
scan the delay $T_V$.  The result of the scan after background
subtraction is shown as the solid circles in Fig.3. The arrow in
the inset shows the location of the delay $T_H$. The solid curve
is a least square Gaussian fit with a visibility of ${\cal V}_3 =
91\%$ and a full width at half height (FWHH) of 185 $\mu m$. Next
we set $T_H$ far away from the peak of the bump as indicated by
the arrow in the inset of Fig.4. We measure again the four-photon
coincidence as a function of $T_V$. The background corrected data
is shown in Fig.4 as the solid circles. The two interference dips
are from the overlap of the V-photon with the two well separated
H-photons, respectively. The solid curve is a least square double
Gaussian fit with a FWHH = 200 $\mu m$ and ${\cal V}_3 = 45\%,
39\%$, respectively. As expected from the prediction of
Ref.\cite{ou}, the single dip in Fig.3 with close to 100\%
visibility corresponds to the indistinguishable case whereas the
double dips in Fig.4 with close to 50\% visibility to the case of
two distinguishable H-photons. The deviations from the exact
predicted values are caused by the less than perfect situations to
be discussed in the following.

There are two origins of imperfection. The first one is from
spatial misalignment of all the fields. This is equivalent to
spatial mode mismatch. In fact, this mode mismatch has been shown
up in the two-photon coincidences (not plotted) between any two of
the ABC detectors. Ideal two-photon interference in three-photon
NOON state projection scheme should have 50\% visibility but the
observed two-photon visibility is 48\% in the case of Fig.3 and is
46\% in the case of Fig.4, which lead to a reduction factor of
$\beta =96\%, 92\%$, respectively.

The second cause is the temporal mode mismatch between the two
pairs of photons generated from two crystals. This type of mode
mismatch between different pairs of photons was encountered in
numerous four-photon interference in parametric down-conversion.
The best way to characterize this mismatch is by a quantity ${\cal
E/A}$ with ${\cal E}$ and ${\cal A}$ defined in Eqs.(2.15,2.16) of
Ref.\cite{orw2} and in general,  we have ${\cal E}\le{\cal A}$.
One extreme value of ${\cal E}/{\cal A}=1$ corresponds to the
situation when the two pairs from parametric down-conversion are
completely overlapping in time and become indistinguishable four
photons. The other extreme case of ${\cal E}/{\cal A}=0$ is for
the situation when the two pairs are well separated and become
distinguishable. A simple application of the theory in
Ref.\cite{orw2} to the situation here gives that
\begin{eqnarray}
&&{\cal V}_3 (T_H<<T_c) = \beta ({\cal A}+3{\cal E})/2({\cal
A}+{\cal E}),\\
&&{\cal V}_3^{(1)} (T_H>>T_c) = \beta /2,\\
&&{\cal V}_3^{(2)} (T_H>>T_c) = \beta {\cal E}/2{\cal
A}.\label{2nd}
\end{eqnarray}
From the measured values of $\beta$ and ${\cal V}_3$, we may
deduce the value of ${\cal E}/{\cal A}$ as
\begin{eqnarray}
&& {\cal E}/{\cal A} = 0.82~~~{\rm from~Eq.(6)},\\
&& {\cal E}/{\cal A} = 0.86~~~{\rm from~Eqs.(7,8)}.
\end{eqnarray}

Obviously, the two dips originate from the overlap of the V-photon
with one of the two H-photons. To understand the difference in the
visibility of the two dips, we plot in Fig.4 the four-photon
coincidence (red diamond) deduced from two-photon coincidence by
the formula
\begin{eqnarray}
&&R_{ABCD}(2\times2)  = (R_{AB}R_{CD} + R_{AC}R_{BD} + \cr&&\hskip
1.5in + R_{AD}R_{BC})/R_0,\label{2x2}
\end{eqnarray}
where $R_0$ is the repetition rate of the pump pulses. The label
$2\times2$ indicates that Eq.(\ref{2x2}) is based on the
assumption that the two pairs are completely independent and the
four-photon coincidence is purely accidental. The appearance of
the lone dip in $R_{ABCD}(2\times2)$ is due to the two-photon
interference of the H and V photons from the same (second)
crystal. The overlap of the first dip in $R_{ABCD}$ with the lone
dip in $R_{ABCD}(2\times2)$ indicates that it is a two-photon
interference effect. So the visibility of this dip does not depend
on whether the two pairs are indistinguishable or not. The
disappearance of the second dip in $R_{ABCD}(2\times2)$ is
consistent with Eq.(\ref{2nd}) because the $2\times 2$ case leads
to ${\cal E}/{\cal A} =0$. So the dependence on ${\cal E}/{\cal
A}$ for the visibility of the second dip in $R_{ABCD}$
[Eq.(\ref{2nd})] indicates that the second dip is due to the
overlap of the H-photon from the first crystal with the V-photon
from the second crystal and the effect indeed requires the
indistinguishability between the two pairs of photons, one from
each crystal.

There is another independent way to measure the value of ${\cal
E}/{\cal A}$, that is, to compare $R_{ABCD}$ and
$R_{ABCD}(2\times2)$ when $T_H=0$ and $T_V=\pm\infty$. For this
purpose, we plot $R_{ABCD}(2\times2)$ (red diamond) and the
corresponding Gaussian fit (red solid curve) in Fig.3. By a
similar analysis that leads to Eqs.(6-8), we find that the ratio
$R_{ABCD}/R_{ABCD}(2\times2)$ at the wings of the scan in Fig.3 is
$1+{\cal E/A}$. From the best fit values in Fig.3, we obtain
${\cal E/A} = 0.81$. This value is consistent with the values in
Eqs.(9,10) which are derived from the visibility in Fig.3 and 4.
The visibility of 48\% for the $R_{ABCD}(2\times2)$ curve in Fig.3
also leads to $\beta = 0.96$ for the $T_H=0$ case, which is
consistent with the two-photon data.

The multi-photon interferometric NOON state projection measurement
scheme is a generalization of the well-known Hong-Ou-Mandel
two-photon interferometer to arbitrary N-photon case. This
statement is based on the facts that (1) this multi-photon
interferometric scheme for $N=2$ is exactly the Hong-Ou-Mandel
two-photon interferometer and (2) the N-photon coincidence is
always zero for any of the state $|k\rangle_H|N-k\rangle_V
(k\ne0,N)$.

\begin{acknowledgments}
This work was funded by National Fundamental Research Program of
China (2001CB309300), the Innovation funds from Chinese Academy of
Sciences, and National Natural Science Foundation of China (Grant
No. 60121503 and No. 10404027)). ZYO is also supported by the US
National Science Foundation under Grant No. 0245421 and No.
0427647.
\end{acknowledgments}

\end{document}